\documentclass[twocolumn,amsmath,amssymb]{revtex4}
\usepackage{graphicx}
\usepackage{dcolumn}
\usepackage{bm}
\usepackage{color}
\begin{document}

\title{Wannier-Stark states of graphene in strong electric field}

\author{Hamed Koochaki Kelardeh$^1$}
\author{Vadym Apalkov$^1$}
\author{Mark I. Stockman$^{1,2,3}$}
\affiliation{
$^1$Department of Physics and Astronomy, Georgia State
University, Atlanta, Georgia 30303, USA\\
$^2$Fakult\"at f\"ur Physik, Ludwig-Maximilians-Universit\"at, Geschwister-Scholl-Platz 1, D-80539 M\"unchen, Germany\\
$^3$Max-Planck-Institut f\"ur Quantenoptik, Hans-Kopfermann-Strasse 1, D-85748 Garching, Germany
}

\date{\today}
\begin{abstract}
We find theoretically energy spectrum of a graphene monolayer in a strong constant electric field using a tight-binding model. Within a single band, we find quantized equidistant energy levels (Wannier-Stark ladder), separated by the Bloch frequency. Singular interband coupling results in mixing of the states of different bands and anticrossing of corresponding levels, which is described analytically near Dirac points and is related to the Pancharatnam-Berry phase. The rate of interband tunneling, which is proportional to the anticrossing gaps in the spectrum, is only inversely proportional to the tunneling distance, in a sharp contrast to conventional solids where this dependence is exponential. This singularity will have major consequences for graphene behavior in strong ultrafast optical fields, in particular, leading to non-adiabaticity of electron excitation dynamics.
%
\end{abstract}
\maketitle

\section{Introduction}

Dynamics of an electron in periodic potential and external electric field is characterized by Bloch oscillations \cite{Bloch_Z_Phys_1929_Functions_Oscillations_in_Crystals}, 
which is a feature of the intraband electron dynamics, and 
Zener tunneling \cite{Zener_Proc_Royal_Soc_1934_Breakdown}, which is related to interband coupling.
The Bloch oscillations occur due to acceleration of an electron by electric field,
which is described by the ``acceleration theorem" in the reciprocal space \cite{Wannier_1959_Book_Solid_State}, and 
subsequent Bragg reflections from the periodic lattice potential at the boundaries of the first Brillouin zone.
The interference of the electron wave packet, following such periodic dynamics in the reciprocal 
space, results in Wannier-Stark (WS) localization of an electron in the coordinate space  \cite{Wannier_1959_Book_Solid_State, Wannier_PR_1960_Wannier_States_in_Strong_Fields}.  
These WS states within a given band are separated by the Bloch oscillation frequency \cite{Bloch_Z_Phys_1929_Functions_Oscillations_in_Crystals}  forming an equidistant WS ladder. 
The Bloch oscillations and the corresponding WS states have been observed experimentally in semiconductor 
superlattices \cite{Bleuse_Bastard_Voisin_PRL_1988_Superlattice_WS_Localization_Theory,
Mendez_et_al_PRL_1988_PhysRevLett.60.2426_WS_Localization_Observation, Feldmann_et_al_PRB_1992_Bloch_Oscillations_Observation_in_Superlattices,
Mendez_Bastard_Phys_Today_1993_WS_Ladders_in_Superlattices, Dekorsy_et_al_SST_1994_Boch_Oscillations}. 
Recently the Bloch oscillations were reportedto to play a major role in high harmonic generation by intense infrared   \cite{Reis_et_al_Nature_Phys_2011_HHG_from_ZnO_Crystal} and terahertz 
 \cite{Huber_et_al_2014_nphoton_2013_349_HHG_in_GaSe} pulses in crystalline solids.

The external electric field not only modifies the intraband electron dynamics, which results in the 
formation of the WS states, but also introduces  
interband coupling of the states of different bands. Such coupling can be described in terms of the Zener 
tunneling resulting in finite widths of the WS levels (resonances) of individual bands \cite{Aldo_1994_Theory_of_Zener_tunneling_and_Wannier_Stark_states_in_semiconductors,
Gluck_1999_Lifetime_of_Wannier_Stark_States,
Gluck_2000_Resonant_tunnelling_of_Wannier_Stark_states,
Gluck_2002_Wannier_Stark_resonances_in_optical_and_semiconductor_superlattices}, or in terms of 
eigenstates of coupled Hamiltonian, which results in mixing of the corresponding WS states of different bands.  
The strongest mixing occurs in the resonance, when the energies of the WS levels of different bands are equal. 
As a function of electric field, at these points the levels exhibit anticrossing behavior. 
In time-dependent electric field, e.g., in the electric field of an optical pulse, passing of these anticrossings defines
 time-dependent electron dynamics. This can be described as an adiabatic formation of WS states of different bands with subsequent passage of the anticrossing points. Depending on relation between 
the anticrossing gap and the rate of change of  electric field, the dynamics of this passage can be adiabatic or diabatic \cite{Apalkov_Stockman_PRB_2012_Strong_Field_Reflection}. Such a description of electron dynamics in time-dependent electric field was successfully used for interpretation of experimental results on interaction of ultrashort intense optical pulses with dielectrics \cite{Schiffrin_at_al_Nature_2012_Current_in_Dielectric, Schultze_et_al_Nature_2012_Controlling_Dielectrics}. 

Description of interaction of time-dependent electric field, e.g., optical pulse, with a solid in 
terms of the dynamics of passage of anticrossing points requires knowledge of both the positions 
of the anticrossing points and the magnitudes of the corresponding anticrossing gaps. These parameters depend 
on the band structure of  the solid and on the strength of the interband coupling. Below we study the properties of 
the WS states of monolayer graphene with potential application to the description of the
interaction of strong optical field with electrons in graphene. 

Graphene monolayer \cite{Novoselov_nat_mater_2007, Electronic_properties_graphene_RMP_2009, graphene_advances_2010} has a honeycomb two-dimensional crystal structure with unique energy dispersion relation. Namely, the low-energy excitations are gapless and are described by the Dirac relativistic massless equation with two Dirac cones. Another important feature of this
relativistic energy dispersion is singularity of the interband 
dipole matrix element between the valence and conduction bands at the Dirac points. In this case, the corresponding interband coupling, introduced by an electric field, is 
strong  near the discrete Dirac points. 

Below in this article we show that, due to this property, 
the stationary Schr\"{o}dinger equation in a constant electric field can be solved exactly within the 
nearest neighbor tight-binding model of graphene for the electric field in the rational crystallographic directions. 
Previously, the WS energy spectra of electrons on a honeycomb lattice were studied in Ref.\ \cite{WS_honeycomb_lattice_PRA_2013} in the tight-binding approximation for both rational and irrational directions of the electric field. 
It was shown that for an electric field in a rational direction, there was the WS localization of the electron wave functions in the field directions while in the normal direction they were delocalized. 


\section{Main Equations}

\begin{figure}
\begin{center}\includegraphics[width=0.48\textwidth]{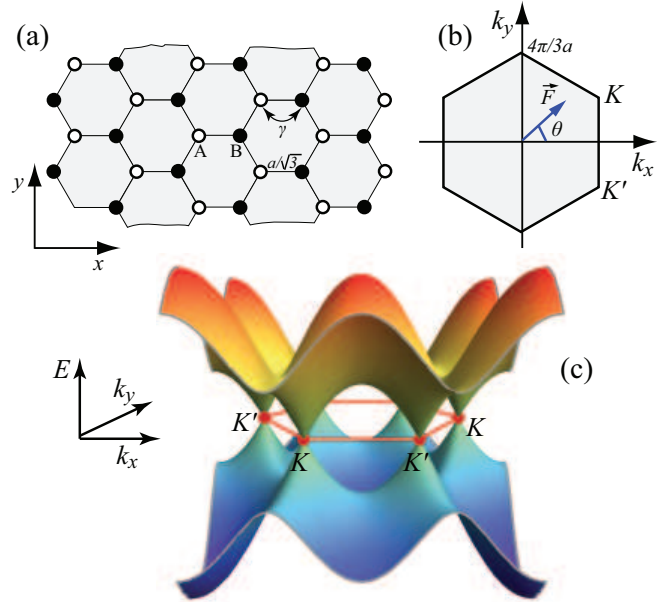}\end{center}
\caption{(a) Honeycomb lattice structure of 2D graphene, which consists of two 
sublattices with atoms labeled by "A" (open circles) and "B" (filled circles), respectively. 
The nearest neighbor coupling with hopping integral $\gamma $ is also shown. 
(b) The first Brillouin zone of graphene. Points $K$ and $K^{\prime }$ are two inequivalent Dirac points, which correspond to two valleys of low energy spectrum of graphene. The direction of electric field is shown by blue line and is characterized by angle $\theta $ relative to the 
$x$ axis.  (c) Energy dispersion of graphene within the nearest neighbor tight-binding model. The $K$ and $K^{\prime }$ Dirac points are labeled. The conduction and the valence bands correspond to positive and negative 
energies, respectively.  
} 
\label{graphene}
\end{figure}

The WS states of an electron in graphene are 
defined as electron states in periodic lattice potential of graphene and in constant 
external electric field. These can be found as solutions of the Schr\"{o}dinger equation, 
\begin{equation}
 {\cal H} \Psi=E\Psi ~,
\label{HPsi}
\end{equation}
where ${\cal H}$ is a single-particle Hamiltonian, which has the  form 
\begin{equation}
{\cal H} = {\cal H}_0 + e \mathbf{F} \mathbf{r} .
\label{Htotal}
\end{equation}
Here ${\cal H}_0$ is a single electron Hamiltonian of graphene, which determines the electron dynamics in periodic lattice potential of graphene, $\mathbf{r} =(x,y)$ is a 2d vector, $e$ is unit charge, and 
$\mathbf{F} = [F\cos\theta , F \sin \theta]$ is the external constant electric field with the magnitude $F$ 
and the direction,  determined by angle $\theta$ relative to the $x$ axis - see Fig.\ \ref{graphene}(b).  

We describe the electron states in graphene within the nearest 
neighbor tight-binding model 
\cite{Graphene_Wallace_PR_1947,Graphene_Weiss_PR_1958,graphene_Dresselhaus_1998,Carbon_nanotubes_2004}
with the tight-binding coupling between the sites of two sublattices "A" and "B" of graphene crystal structure - 
see Fig.\ \ref{graphene}(a). Such a model describes both the conduction and valence bands of graphene 
and captures the properties of the Dirac points. In the reciprocal space, the tight-binding Hamiltonian 
${\cal H}_0$ can be represented by a $2\times 2$ matrix of the form \cite{Graphene_Wallace_PR_1947,Graphene_Weiss_PR_1958}
\begin{equation}
{\cal H}_0 =  
\left(
\begin{array}{cc}
0  &  \gamma  f(\mathbf{k}) \\
\gamma f^{*}(\mathbf{k})  & 0
\end{array}
\right)   ,
\end{equation}
where $\gamma=-3.03$ eV is the hopping integral and 
\begin{equation}
f(\mathbf{k}) = \exp \left( i \frac{a k_x} {\sqrt{3}}\right) + 2  \exp \left( -i \frac{a k_x} {2 \sqrt{3}}\right) 
 \cos \left( \frac{ak_y }{2} \right).
\end{equation}
Here $a=2.46 \mathrm{\AA }$ is a lattice constant.  
The energy spectrum of Hamiltonian ${\cal H}_0$ consists of conduction band ($\pi^*$ or anti-bonding band) and valence bands ($\pi$ or bonding band) with the energy 
dispersion $E_{c}(\mathbf{k})= -\gamma |f(\mathbf{k})|$ (conduction band) and $E_{v}(\mathbf{k})= \gamma |f(\mathbf{k})|$ (valence band). This energy dispersion is shown in 
Fig.\ \ref{graphene}(c). It consists of two inequivalent sets of three Dirac points (and cones) $\mathbf{K}$ and $\mathbf{K}^{\prime }$.
The corresponding wave functions of the conduction and valence 
bands are, respectively,
\begin{equation}
\Psi^{(c)}_{\mathbf{k}} (\mathbf{r} ) = \frac{e^{i \mathbf{k} \mathbf{r}}}{\sqrt{2}}
\left( 
\begin{array}{c}
1 \\
e^{i \varphi_k }
\end{array}
\right)  ~,
\label{functionV}
\end{equation}
and  
\begin{equation}
\Psi^{(v)}_{\mathbf{k}} (\mathbf{r} ) = \frac{e^{i \mathbf{k} \mathbf{r}}}{\sqrt{2}}
\left( 
\begin{array}{c}
-1 \\
e^{i \varphi_k }
\end{array}
\right)  ~,
\label{functionC}
\end{equation}
where we denote $ f(\mathbf{k}) = |f(\mathbf{k}) | e^{i\varphi _k}$. 
The wave functions $\Psi^{(c)}_{\mathbf{k}}$ and $\Psi^{(v)}_{\mathbf{k}}$  have two
components corresponding to two sublattices A and B.

\begin{figure}
\begin{center}\includegraphics[width=0.45\textwidth]{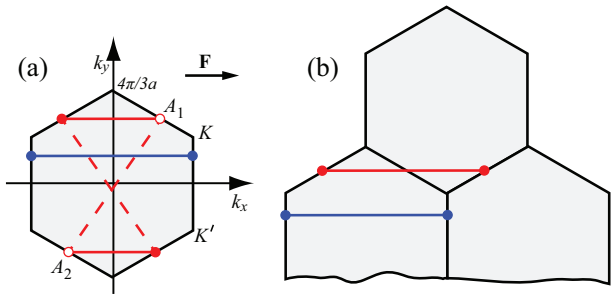}\end{center}
\caption{Lines of coupled states in reciprocal space.   
The electron states of the reciprocal space, which are coupled by a constant electric 
field parallel to the $x$ axis, are shown by solid lines of two different colors (red and blue), where different colors 
correspond to two different values of  $k_y$. 
(a) The coupled states are shown in the first Brillouin 
zone. The equivalent points (at the edges of the Brillouin
zone) are shown by the same type of points, i.e. solid red points or open red points. The equivalent points are 
connected by a vector of reciprocal lattice. (b) The coupled states are shown in the whole reciprocal space. The first Brillouin zones, localized at different points of the reciprocal lattice, are also shown. The equivalent points, which are connected by a vector of reciprocal lattice, are 
shown by the same type of points, e.g., two red points are 
equivalent.  
} 
\label{grapheneFx}
\end{figure}

Taking the eigenfunctions $\Psi^{(v)}_{\mathbf{k}}$ and $\Psi^{(c)}_{\mathbf{k}}$ of 
Hamiltonian ${\cal H}_0$ as the basis, we express the general solution of Schr\"{o}dinger equation (\ref{HPsi}) in the form 
\begin{equation}
\Psi (\mathbf{r}) = \sum_{\mathbf{k}} \left[\phi_v(\mathbf{k}) \Psi^{(v)}_{\mathbf{k}} (\mathbf{r}) +
                                     \phi_c(\mathbf{k}) \Psi^{(c)}_{\mathbf{k}} (\mathbf{r})      \right]~.
\label{PsiVC}
\end{equation}
Expansion coefficients $\phi_v(\mathbf{k})$ and $\phi_c(\mathbf{k})$ satisfy the following 
eigenvalue equations (see the Appendix).
\begin{equation}
E \phi_v(\mathbf{k})  =  E_{v}(\mathbf{k})\phi_v(\mathbf{k})  + 
i e \mathbf{F} \frac{\partial\phi_v(\mathbf{k})}{\partial {\mathbf{k}}}   +  \mathbf{F} \mathbf{D}(\mathbf{k} )  \phi_c(\mathbf{k})~,          
\label{Eqs1}  
\end{equation}
\begin{equation}
E \phi_c(\mathbf{k})  =  E_{c}(\mathbf{k})\phi_c(\mathbf{k})  + 
i e \mathbf{F} \frac{\partial\phi_c(\mathbf{k})}{\partial {\mathbf{k}}}   +  \mathbf{F} \mathbf{D}(\mathbf{k} )    \phi_v(\mathbf{k})~,
\label{Eqs2}
\end{equation}
where $\mathbf{D}(\mathbf{k} ) =[D_x(\mathbf{k}),D_y(\mathbf{k})]  $ is the dipole matrix element between 
the conduction and valence band states with the wave vector $\mathbf{k}$, i.e.,
\begin{equation}
\mathbf{D}(\mathbf{k} ) = \left\langle \Psi^{(c)}_{\mathbf{k} }\right|  e\mathbf{r} 
\left| \Psi^{(v)}_{\mathbf{k} }\right\rangle=\frac{e}{2}\frac{\partial\varphi_\mathbf{k}}{\partial\mathbf{k}}~ .
\label{dipole}
\end{equation}
Substituting conduction and valence band wave functions (\ref{functionV}) and 
(\ref{functionC}) into Eq.\ (\ref{dipole}), we obtain the following 
expressions for the interband dipole matrix elements:
\begin{equation}
D_x(\mathbf{k})  = \frac{e a}{2\sqrt{3}}  \frac{1+ \cos \left( \frac{ak_y}{2}  \right) 
       \left[ \cos \left(\frac{3ak_x}{2\sqrt{3} } \right) - 2\cos \left(\frac{ak_y}{2}  \right)   \right]  }
         {1+4 \cos \left( \frac{ak_y}{2}  \right) 
       \left[ \cos \left(\frac{3ak_x}{2\sqrt{3} } \right) + \cos \left(\frac{ak_y}{2}  \right)   \right]  } ~,
\label{Dx}
\end{equation}
and 
\begin{equation}
D_y(\mathbf{k})  = \frac{e a}{2}  \frac{\sin \left( \frac{ak_y}{2}  \right)  
      \sin \left(\frac{3ak_x}{2\sqrt{3} } \right) }
         {1+4 \cos \left( \frac{ak_y}{2}  \right) 
       \left[ \cos \left(\frac{3ak_x}{2\sqrt{3} } \right) + \cos \left(\frac{ak_y}{2}  \right)   \right]  }~. 
\label{Dy}
\end{equation}
A solution, $\phi_v(\mathbf{k})$ and $\phi_c(\mathbf{k})$, of Eqs. (\ref{Eqs1})-(\ref{Eqs2}) should satisfy periodic boundary condition in the reciprocal space with the periodicity of the reciprocal lattice. From this condition, we obtain the WS energy spectrum -- Sec.\ \ref{Results} below.

Equations (\ref{Eqs1})-(\ref{Eqs2}) constitute a system of the first order differential equations, where  
a constant electric field introduces both interband and intraband coupling of the electron 
states. The interband coupling is realized only between the states with the same wave vector, while the 
intraband coupling occurs only between the states laying in the reciprocal space along a trajectory 
determined by the direction of electric field.
These trajectories can be identified 
by considering electron dynamics in a reciprocal space in a constant electric field. If an electron is initially at some point $\mathbf{k}$ of the reciprocal space and a constant electric field is 
applied, then this electron will drift along the direction of the electric field following the acceleration theorem, $\hbar d\mathbf{k}/dt = e\mathbf{F}$,  experiencing Bragg scattering at the 
boundaries of the Brillouin zone. Then the corresponding electron trajectory in the reciprocal space determines the line of coupled states. 

The intraband-coupled states can be described by considering the states either in the first  Brillouin 
zone only or in the entire reciprocal space. In either case, the equivalence of the points connected 
by a vector of reciprocal lattice should be taken into account. Such equivalence determines the 
periodic boundary conditions in the reciprocal space, from which the energy spectrum can be obtained. 

First, we assume that the electric field is parallel to the $x$ axis. In this case, the lines of coupled states 
are also parallel to the $x$ axis and are parametrized by the  $y$ component of the wave vector, $k_y$. 
In Fig.\ \ref{grapheneFx} the states coupled by this electric field are shown  in the first Brillouin zone [Fig.\ \ref{grapheneFx}(a)] and in the extended reciprocal space [Fig.\ \ref{grapheneFx}(b)]. 
In the first Brillouin zone, we need to take into account 
equivalence of the points connected by a vector of the reciprocal lattice, e.g., points $A_1$ and $A_2$ 
are equivalent. In Fig.\ \ref{grapheneFx}, two sets of coupled states (lines) corresponding to different values of 
$k_y$ are shown. 
If  $k_y<2\pi /a$ then the 
typical line of coupled states is shown by the blue solid line in Fig.\ \ref{grapheneFx}. The 
solid blue points at the ends of the line are coupled by a vector of reciprocal lattice, which determines 
the periodic boundary conditions for the wave functions $\phi_v (\mathbf{k})$ and $\phi_c (\mathbf{k})$, i.e., 
$\phi_v (-2\pi/a\sqrt{3},k_y)=\phi_v (2\pi/a\sqrt{3},k_y)$ and $\phi_c (-2\pi/a\sqrt{3},k_y)=\phi_c (2\pi/a\sqrt{3},k_y)$. From these conditions, the energy spectrum is obtained. 

If $k_y > 2\pi/3a$, then the line of coupled states in the first Brillouin zone consists of two 
line segments, which are shown by red solid lines in Fig.\ \ref{grapheneFx}(a). These line segments  
have  two sets of equivalent points: solid red points and open red points. 
The points in each set are connected by the corresponding vector of the reciprocal lattice.

In the extended reciprocal space, a part of which is shown in Fig.\ \ref{grapheneFx}(b), the 
lines, which describe the coupled states, are 
straight lines for both $k_y < 2\pi/3a$ and $k_y > 2\pi/3a$. For the case $k_y > 2\pi/3a$, the line of coupled 
states is located in two Brillouin zones -- see Fig.\ \ref{grapheneFx}(b). For both the red and 
blue lines, the end points are connected by the same vector of reciprocal lattice, $\mathbf{G} = (4\pi/a\sqrt{3},0)$, 
which makes the end points equivalent  and introduces periodic boundary conditions for the system of equations (\ref{Eqs1})-(\ref{Eqs2}). 

\section{Results and Discussion}
\label{Results}

\subsection{Wannier-Stark levels of a single band}

Without interband coupling, i.e., for $\mathbf{D} = 0$, Eqs.\ (\ref{Eqs1})-(\ref{Eqs2}) become decoupled. For a single band, e.g., valence band, Eq.\ (\ref{Eqs1}) becomes
\begin{equation}
E \phi_v(\mathbf{k})  =  E_{v}(\mathbf{k})\phi_v(\mathbf{k})  + 
i e F \frac{d\phi_v(\mathbf{k})}{d {k_x}},  
\label{EqV}
\end{equation}
where the electric field is parallel to the $x$ axis. 
Solution of the first order differential equation (\ref{EqV}) 
has the form 
\begin{eqnarray}
\phi_v^{(0)}(\mathbf{k}) =  \frac{1}{\sqrt{2k_0}} 
\exp \Bigg[ && -\frac{i}{eF} \Bigg( E(k_x+k_0) -    
\nonumber \\
&& 
 \int_{-k_0}^{k_x} E_{v}(k^{\prime },k_y) dk^{\prime } \Bigg) \Bigg],
 \label{WSV1}
\end{eqnarray}
where we introduced a notation, $k_0 = 2\pi/\left(a\sqrt{3}\right)$.
From the periodicity of the wave function, 
$\phi_v(-k_0,k_y)=\phi_v(k_0,k_y)$, we obtain the 
WS energy spectrum as
\begin{equation}
E^{WS}_{v,n} = E_{v,0}(k_y) + \hbar \omega_B n,
\label{EWS}
\end{equation}
where $n$ is an integer, and the band offset, $E_{v,0}(k_y)$, is 
\begin{equation}
E_{v,0}(k_y) = \frac{1}{2k_0} \int_{-k_0}^{k_0} E_{v}(k^{\prime },k_y) dk^{\prime }.
\label{BandOffV}
\end{equation}
 The Bloch frequency $\omega_B$ in Eq.\ (\ref{EWS})  is defined as 
\begin{equation}
\omega_B = \textcolor{black}{\pi \frac{eF}{\hbar k_0 }.}
\label{omegaBB}
\end{equation}
The energy spectrum of Eq.\ (\ref{EWS}) forms the WS ladder 
with equidistant energy levels. 

For the conduction band, the energy spectrum has a similar form, 
\begin{equation}
E^{WS}_{c,n} = E_{c,0}(k_y) + \hbar \omega_B n,
\label{EWScb}
\end{equation}
with the corresponding band offset 
\begin{equation}
E_{c,0}(k_y) = \frac{1}{2k_0} \int_{-k_0}^{k_0} E_{c}(k^{\prime },k_y) dk^{\prime }.
\label{BandOffC}
\end{equation}
For the tight-binding model, introduced above, there is a relation  $ E_{c,0}(k_y) = - E_{v,0}(k_y)$.
The wave functions of the WS levels of the conduction band are  
\begin{eqnarray}
\phi_c^{(0)}(\mathbf{k}) =  \frac{1}{\sqrt{2k_0}} 
\exp \Bigg[ && -\frac{i}{eF} \Bigg( E(k_x+k_0) -    
\nonumber \\
&& 
 \int_{-k_0}^{k_x} E_{c}(k^{\prime },k_y) dk^{\prime } \Bigg) \Bigg],
\label{WSC1}
\end{eqnarray}
In the coordinate space, the WS levels are localized and the integer index $n$ 
determines the center of localization.

\subsection{Wannier-Stark states of two-band model: Analytical results} 
\label{Model1DP}

\subsubsection{Energy spectrum}

\begin{figure}
\begin{center}\includegraphics[width=0.45\textwidth]{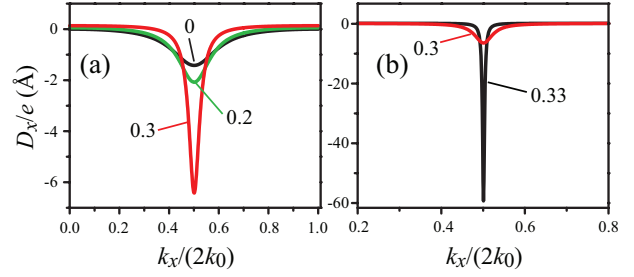}\end{center}
\caption{Dipole matrix element $D_x$ as a function of  $k_x$ for different values of $k_y$. 
The wave vector $k_x$ is measured in units of 
$2k_0$, where $k_0 = 2\pi/a\sqrt{3}$. The Dirac point is at $k_x = k_0$ and $k_y=k_0/\sqrt{3}=(1/3) (2\pi/a)$. 
The numbers near the lines are the values of $k_y$ in units of $(2\pi/a)$. Panels (a) and (b) differ by the vertical scale. 
} 
\label{FigZ}
\end{figure}

The interband coupling, determined by dipole matrix elements $\mathbf{D}(\mathbf{k})$, has a
strong dependence on wave vector $\mathbf{k}$. Near the Dirac points ($K$ and $K^{\prime }$ points in 
Fig.\ \ref{graphene}), the dipole matrix elements have sharp peaks. Dependence of the dipole matrix element, $D_x$, on 
the wave vector, $k_x$, for different values of $k_y$ is shown in Fig.\ \ref{FigZ}. The $K$ Dirac point is at $\mathbf{k}=\mathbf{K} \equiv (K_x, K_y)=
(2\pi/a)(1/\sqrt{3},1/3)$, i.e. it corresponds to $k_y = K_y = 
(1/3)(2\pi/a)=k_0/\sqrt{3}$ and $k_x = K_x = (2\pi/a\sqrt{3})=k_0$. Away from 
the Dirac point, i.e., when $\textcolor{black}{\left|k_y\right|}\ll K_y$ 
 [see Fig.\ \ref{FigZ}(a)], the dipole matrix element, $|D_x|$,  has a broad maximum near $k_x = k_0$. With increasing $k_y$,
the maximum becomes more pronounced. Near the Dirac point 
[see, e.g., the case of 
$k_y = 0.33 (2\pi/a)$ in Fig.\ \ref{FigZ}(b)], the dipole matrix element, $|D_x|$, has a sharp peak at $k_x = K_x=k_0$. Near this peak, 
the dipole matrix element, $D_x(k_x,k_y)$, behaves as 
\begin{equation}
D_x(k_x,k_y) \approx \frac{3ea}{4 \pi} \left[     
      -\frac{1}{\delta_y}  + 
        \frac{3}{\delta_y^3} \delta_x^2 \right],
\label{Dapprox}
\end{equation}
where $\delta_y = (k_y - K_y)/K_y$ and $\delta_x = (k_x - K_x)/K_x$.
Thus, for a given $k_y$, the maximum value of the dipole matrix element is $(3ea/4 \pi) \left[ K_y/(k_y-K_y) \right] $, diverging at $k_y \rightarrow K_y$.

Although the shape of $D_x(k_x,k_y)$ as a function of $k_x$ 
depends on the value of the $y$ component of the wave vector, 
$k_y$, the net interband coupling, which can be characterized 
by the integral 
\begin{eqnarray}
\frac{1}{e}D_{x}^{(net)} (k_y) &=&\frac{1}{e} \int_{-k_0}^{k_0} D_x(k_x,k_y) dk_x=
\nonumber\\
&&\frac{\varphi_\mathbf{k}}{2}\Big\vert_{k_x=-k_0}^{k_x=k_0}= - \frac{\pi}{3}~,
\label{Dnet}
\end{eqnarray} 
which does not depend on $k_y$. The interband transitional dipole, $D_{x}^{(net)} (k_y)$, is determined by the Pancharatnam-Berry phase, $\left(\varphi_\mathbf{k}/2\right)\Big\vert_{k_x=-k_0}^{k_x=k_0}$ \cite{Pancharatnam_1956_PIAS_Geometric_Phase, Berry_Phase_Proc_Royal_Soc_1984}, as characteristic of dielectric responses of crystalline solids  \cite{Zak_PhysRevLett.62_1989_Berry_Phase_in_Crystals, Resta_RMP_1994_Berry_Phase, Xiao_Niu_RevModPhys.82_2010_Berry_Phase_in_Electronic_Properties}. This also suggests that Eq.\ (\ref{Dnet}), as defined by the symmetry of the system, is more general than the tight-binding model, in which the specific calculations are made.

A strong dependence of the dipole matrix element on $k_x$ near the Dirac point, which is illustrated in Fig.\ \ref{FigZ} and is 
supported by Eq.\ (\ref{Dapprox}), can be approximated by the  $\delta$-function, i.e. 
\begin{equation}
D_x(k_x,k_y) = e\Lambda_0 \delta(k_x - k_0).
\label{Ddelta}
\end{equation}
Here strength $\Lambda _0$ of the $\delta$-function  
is determined by the condition that the net dipole coupling 
in Eqs.\ (\ref{Dnet}) and (\ref{Ddelta}) is the 
same, which yields $\Lambda_0 = -\pi/3$.

For the $\delta$-function profile of the dipole matrix elements, the system of equations (\ref{Eqs1})-(\ref{Eqs2}) 
can be solved analytically. Such solution can be obtained as follows. We are looking for a solution of the system of equation 
(\ref{Eqs1})-(\ref{Eqs2}) within a line segment $0\leq k_x <2k_0$ with the periodical boundary conditions at the ends. (Here, it is convenient to 
consider interval $0\leq k_x <2k_0$ and not  interval $-k_0\leq k_x <k_0$ introduced before.) The dipole matrix element is 
non-zero only at $k_x = k_0$. Then, for $0\leq k_x <k_0$ and  $k_0< k_x <2k_0$, there is no interband coupling between 
the valence and conduction bands. Within these intervals, the general solution of the system (\ref{Eqs1})-(\ref{Eqs2}) acquires the form for  $0\leq k_x <k_0$,
\begin{eqnarray}
\phi_v(\mathbf{k}) &=& A_1 \exp{}\Bigg[-\frac{i}{eF} \bigg( Ek_x -
\nonumber\\
& &\int_{0}^{k_x} E_{v}(k^{\prime },k_y) dk^{\prime } \bigg)\Bigg] ~, 
\label{phiVD1} \\
\phi_c(\mathbf{k}) &=& A_2 \exp{}\Bigg[-\frac{i}{eF} \bigg( E k_x -
\nonumber\\
& &\int_{0}^{k_x} E_{c}(k^{\prime },k_y) dk^{\prime } \bigg)\Bigg]  ~,
\label{phiCD1}
\end{eqnarray}  
and \textcolor{black}{the same form with different coefficients} for $k_0< k_x <2k_0$,
\begin{eqnarray}          
\phi_v(\mathbf{k}) &=& B_1 \exp{}\Bigg[-\frac{i}{eF} \bigg( Ek_x -
\nonumber\\
& &\int_{0}^{k_x} E_{v}(k^{\prime },k_y) dk^{\prime } \bigg)\Bigg] ~, 
\label{phiVD2}\\
\phi_c(\mathbf{k}) &=& B_2 \exp{}\Bigg[-\frac{i}{eF} \bigg( E k_x -
\nonumber\\
& &\int_{0}^{k_x} E_{c}(k^{\prime },k_y) dk^{\prime } \bigg)\Bigg]  ~,
\label{phiCD2}
\end{eqnarray}
 where $A_1$, $A_2$, $B_1$, and $B_2$ are constants. 

At point $k_x = k_0$, the $\delta$-function dependence of dipole matrix element (\ref{Ddelta}) introduces the following 
relation between the values of the wave function at $k_x = k_0-0$ and $k_x = k_0 + 0$:
\begin{eqnarray}
& & \phi_v|_{k_0 + 0} = -i \phi_c|_{k_0 - 0} \sin \Lambda_0 + \phi_v|_{k_0 - 0} \cos \Lambda_0    \label{relation1} \\
& & \phi_c|_{k_0 + 0} = \phi_c|_{k_0 - 0} \cos \Lambda_0 -i \phi_v|_{k_0 - 0} \sin \Lambda_0   \label{relation2}
\end{eqnarray}
Thus, the $\delta$-function coupling results in rotation of a pseudospin, which is associated with two components of the wave function, by a 
finite angle $\Lambda_0$. 

Substituting expressions (\ref{phiVD1})-(\ref{phiCD2}) into relations (\ref{relation1})-(\ref{relation2}) and taking into 
account the periodic boundary conditions, we obtain an equation for the energy 
spectrum of the WS states,
\begin{equation}
\cos \left(\frac{2k_0}{eF} E    \right) = \cos (\Lambda_0) \cos \left( \frac{2k_0}{eF}  E_{c,0}(k_y)   \right),
\label{WSdeltaEq}
\end{equation}
where we took into account relation $E_{c,0}= -E_{v,0}$, which is valid within the tight-binding model introduced above. 
The solution of Eq.\ (\ref{WSdeltaEq}) is parametrized by an integer number $n$; it describes the WS-state energies and has the form 
\begin{eqnarray}
&&E_{n}^{(\pm)} = \pm \frac{eF}{2k_0}\times
\nonumber\\
&& \left\{ 
\cos^{-1} \left[ \cos \Lambda_0 \cos \left( \frac{2k_0}{eF}  E_{c,0}(k_y)   \right)  \right]
   + 2\pi n \right\}.
\label{EnWS}
\end{eqnarray}
Here the $\pm$ signs correspond to the the conduction ($c$) and valence ($b$) bands, respectively. 

It is convenient to rewrite Eq.\  (\ref{EnWS})  in dimensionless energy variables normalized to the Bloch frequency,
$\varepsilon_n^{(\pm)} = E_n^{(\pm)}/\hbar \omega_B = E_n^{(\pm)} \textcolor{black}{\left(k_0/(\pi eF)\right)}$ and 
$\varepsilon_{c,0} =  E_{c,0}(k_y)/\hbar \omega_B =  E_{c,0}\textcolor{black}{\left(k_0/(\pi eF)\right)}$ as
\begin{equation}
\varepsilon_n^{(\pm)} = \pm (2\pi)^{-1}\cos^{-1} \left[ \cos \Lambda_0 \cos  \left(2\pi\varepsilon_{c,0}\right)    \right]
   + \textcolor{black}n .
\label{EnWSdim}
\end{equation}

\begin{figure}
\begin{center}\includegraphics[width=0.4\textwidth]{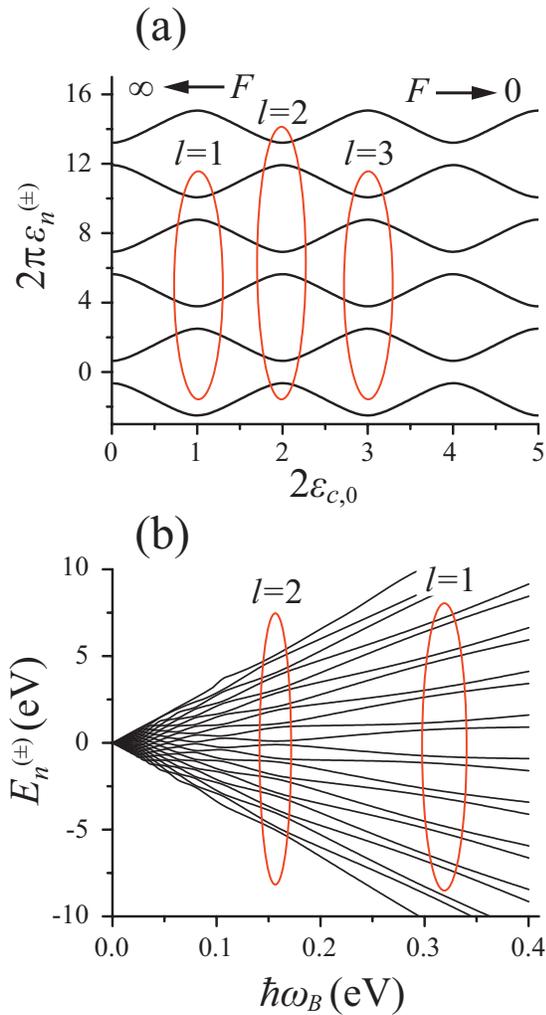}\end{center}
\caption{(a) Dimensionless energies $\varepsilon_n^{(\pm)}$ of the WS states, calculated from 
Eq.\ (\ref{EnWSdim}), 
as a function of dimensionless parameter $\varepsilon_{c,0}$ for different values of integer number $n$. Parameter 
$\Lambda_0$ is $\Lambda_0=0.6$. Different types of anticrossing points are labeled by integer parameter $l$. With 
increasing electric field, the last anticrossing point corresponds to $l=1$ and occurs at $\varepsilon_{c,0}=\pi$.
(b) The energies $E_n^{(\pm)}$ of the WS states, calculated from Eq.\ (\ref{EnWS}), as a function 
of Bloch frequency, $\hbar \omega_B$, which is proportional to electric field. The anticrossing points, corresponding 
to $l=1$ and $l=2$, are marked by red lines. The parameter $\Lambda_0$ is $\Lambda_0=0.6$, and $E_{c,0} =1$ eV. 
} 
\label{FigModel}
\end{figure}

The corresponding dimensionless energy spectrum is shown in Fig.\ \ref{FigModel}(a). The anticrossing points of the energy levels 
can be clearly identified. These points are the anticrossings of the WS ladders of the conduction and valence 
 bands -- see  Fig.\ \ref{FigModel}(b) -- corresponding to the interband Zener tunneling \cite{Zener_Proc_Royal_Soc_1934_Breakdown}. This interband coupling (Zener tunneling) makes the initial WS states of isolated bands to be non-stationary (metastable) but causes the formation of new, stationary states of the coupled bands that we consider in this article.

The anticrossing points
can be labeled by an integer number $l=1,2,\dots$. which has meaning of the number of unit cells through which the Zener tunneling occurs. In dimensionless variables, the positions of the anticrossing points are 
\begin{equation}
\varepsilon_{c,0}^{(l)} = l/2~,
\label{pointsDim}
\end{equation}
or, in terms of the electric field, the anticrossing points are at 
\begin{equation}
F^{(l)} =    \frac{2  k_0}{e\pi l} E_{c,0}.
\label{Fexact}
\end{equation}

The positions of the anticrossing points can be also estimated from the expressions (\ref{EWS}), (\ref{EWScb}) for the 
energies of the WS states of uncoupled conduction and valence bands. For uncoupled bands, the anticrossing points are 
determined by an equation $E^{WS}_{c,n_c} = E^{WS}_{v,n_v}$, from which one can derive the positions of the 
anticrossing points at
\begin{equation}
F^{(l)}_{\mbox{uncoupled}} = \frac{2  k_0}{\pi e l} E_{c,0},
\label{Funcoupled}
\end{equation}
where $l = n_c - n_v$. 
Comparing exact expression (\ref{Fexact}) with approximation (\ref{Funcoupled}),  
we can conclude that the interband coupling for $k_y$ in the vicinity of the Dirac point eliminates field-induced renormalization of an anticrossing position,
\begin{equation}
\textcolor{black}{F^{(l)} = F^{(l)}_{\mbox{uncoupled}}~.}
\end{equation}
At the same time, for $k_y$ far from the Dirac point,
the interband coupling shifts the anticrossing points to the higher 
values of electric field similar to ordinary 3d solids \cite{Apalkov_Stockman_PRB_2012_Strong_Field_Reflection} (see also below in Sec.\ \ref{WS_2_band}). 

In the dimensionless units, the anticrossing gaps are the same for all anticrossing points [cf.\ Fig.\ \ref{FigModel}(a)]. 
The value of the dimensionless gap, $\Delta_g/\hbar \omega_B$, can be found as the difference between the corresponding energy levels, $\Delta_g/\hbar \omega_B = \varepsilon_1^{(-)} - \varepsilon_0^{(+)}$, calculated at a point $\varepsilon_{c,0} = 1/2$. This way, we find 
\begin{equation}
\Delta_g/\hbar \omega_B =  \Lambda_0/\pi~.
\end{equation}
In the original units, the anticrossing gap corresponding to the anticrossing point with index $l$ [see
  Eq.\ (\ref{pointsDim})] takes the form 
\begin{equation}
\Delta_g^{(l)} = \frac{2\Lambda_0}{l\pi} E_{c,0}.
\label{Dgap}
\end{equation}
Such weak dependence of the anticrossing gap on index  $l$ is a unique feature of graphene's
unconventional relativistic-like low-energy dispersion relation. This behavior is quite different from that of conventional solids, e.g., dielectrics, for which the anticrossing gaps are exponentially decreasing with $l$. 

The physical meaning of $l$ is that the value of 
$al$ is the distance between 
the localized WS states of the conduction and valence bands. Then, the anticrossing gap with index $l$ 
is determined by a coupling of the WS states of the conduction and valence bands separated by 
spatial distance $al$ and is equal to the rate of Zener tunneling \cite{Zener_Proc_Royal_Soc_1934_Breakdown} between these bands through $l$ unit cells in space.
 For graphene, such coupling has a long range in the direct space due to the strongly 
localized $\delta$-function profile of the dipole matrix elements in the reciprocal space. Such a long-range tunneling results in a weak dependence of the anticrossing gap on distance $l$. 

The $\delta$-function profile of the dipole matrix elements in graphene is an approximation, used above to obtain analytical solution 
of the problem. The exact dipole matrix element $\mathbf{D}(\mathbf{k})$ has a finite small width $w_D$ in the 
reciprocal space (see Fig.\ \ref{FigZ}), where  $w_D$ depends on $k_y$. Such a finite width introduces a 
cutoff both in the long-range coupling of the WS states of different bands and in the weak dependence of the 
anticrossing gap on $l$. Namely, the anticrossing gap $\Delta_g^{(l)}$ has the weak, $l^{-1}$, dependence 
on $l$ for $l\lesssim l_c= (w_D a)^{-1}$; for $l\gg l_c$, the anticrossing gap becomes exponentially small with $l$.

Since the dimensionless parameter $\varepsilon_{c,0} $ is inversely proportional to electric field, then in 
the energy spectrum, considered as a function of electric field, the anticrossing point with 
index $l=1$ is the last anticrossing point [see Fig.\ \ref{FigModel}(a)]. 
In Fig.\ \ref{FigModel}(b) the energy spectrum, calculated from 
Eq.\ (\ref{EnWS}), is shown as a function of electric field. The anticrossing points with indexes $l=1$ and $l=2$ are 
marked. The corresponding anticrossing gaps are given by Eq.\ (\ref{Dgap}). The last anticrossing points with index $l=1$ has the largest anticrossing gap, $\Delta_g^{(1)} = 2\Lambda_0 E_{c,0}/\pi$.

For graphene, within the tight-binding model introduced above, parameter $\Lambda_0$, calculated at $k_y =k_{y,0} = 2\pi/3a$, is $|\Lambda_0| =\pi/3 \approx 1.05$. For this value of $k_y=k_{y,0}$, the energy dispersion is 
\begin{equation}
E_c (k_x,k_{y,0}) = -2\gamma \cos \left( \frac{\sqrt{3} a k_x}{ 4 } \right).  
\label{EcDirac} 
\end{equation}
Then the band offset of the conduction band, defined by Eq.\ (\ref{BandOffC}), is 
\begin{equation}
E_{c,0} (k_{y,0}) = -\frac{4\gamma }{\pi} \approx 3.86 ~\mathrm{eV}.
\end{equation}
 For these values of $\Lambda_0$ and $E_{c,0}$, we obtain from Eqs.\ (\ref{Fexact}) and (\ref{Dgap})
the positions of the anticrossing points and the corresponding anticrossing gaps
\begin{eqnarray}
& & F^{(l)} = \frac{8k_0 |\gamma|}{e\pi^2 l}  \approx    \frac{3.59}{l} ~\frac{\mathrm{V}}{\mathrm{\AA}}~ ,  
\label{FexactGr} \\
& & \Delta_g^{(l)}= \frac{8 |\gamma |}{3\pi l} \approx \frac{2.54}{l} ~\mathrm{eV}~. 
\label{GapexactGr}
\end{eqnarray}
The anticrossing at $l=1$ is  the last one occuring at the maximum electric field of $3.59$ V/\AA. The anticrossing gap at this point is $2.54$ eV.

\subsubsection{Wave functions}

The wave functions of the WS states of the two-band graphene model have two components, $ \phi_v(\mathbf{k})$ and 
$\phi_c(\mathbf{k})$, which give the amplitudes for an electron to be in the valence and conduction band, respectively. These functions, $\phi_v(\mathbf{k})$ and $\phi_c(\mathbf{k})$, are determined by Eqs.\ (\ref{phiVD1})- 
(\ref{phiCD2}) where the unknown coefficients $A_1$, $A_2$, $B_1$, and $B_2$ can be found from the boundary conditions (\ref{relation1})-(\ref{relation2}). At a given energy of the WS state $E$, they
have the following form 
\begin{eqnarray}
& & A_2 = A_1 \exp{}\left\{\frac{-i}{eF}\int_{0}^{2k_0} \left[E_{c}(k^{\prime },k_y)-E\right] dk^{\prime }\right\},~~
%
\label{A2} \\
& & B_1 = i\frac{ A_2 - A_1 \cos(\Lambda_0)}{\sin(\Lambda_0)}~,~~
\\
& & B_2 = B_1 \cos(\Lambda_0) - i A_1 \sin(\Lambda_0)~.
\end{eqnarray}  
Here coefficient $A_1$ can be found from the normalization condition. 
The wave functions, $ \phi_v(\mathbf{k}),\phi_c(\mathbf{k})$, determine the electron amplitudes in the reciprocal 
space. The corresponding wave functions in the direct coordinate space are determined by a Fourier transform, 
\begin{eqnarray}
& & \tilde{\phi}_v(x,k_y) = \int dx \phi_v(k_x,k_y) e^{ik_x x},
 \label{phiVx} \\
& & \tilde{\phi}_c(x,k_y) = \int dx \phi_c(k_x,k_y) e^{ik_x x}, \label{phiCx}
\end{eqnarray}  
where we consider the spatial dependence of the wave function along axis $x$ only, i.e., along 
the direction of the electric field. In this case, the $y$ component of the wave vector, $k_y$, 
should be considered as a parameter. 

Without interband coupling, i.e., for $\Lambda_0 = 0$, and for $k_y = k_{y,0}$, 
the WS wave functions for a given band, e.g., conduction band, can be expressed in terms of the 
Bessel functions, 
\begin{equation}
\tilde{\phi}_c(x,k_y) \propto J_{\left| \frac{4}{\sqrt{3} a} \left( x - \frac{E}{e F}\right) \right|} 
\left( \frac{\gamma}{\hbar \pi \omega_B}  \right),
\label{WSsingle}
\end{equation}
where $J_n(z)$ is the Bessel function of order $n$, and the Bloch frequency is given by Eq.\ (\ref{omegaBB}). 
Such analytical expression is obtained for energy dispersion (\ref{EcDirac}). Wave function (\ref{WSsingle}) is localized in the $x$-space at a coordinate point $x = E/eF$, which is proportional to the energy of the WS state. 

The interband coupling, $\Lambda_0$, results in mixing 
of the wave functions of different (conduction and valence) bands. 
The mixing is strongest at the anticrossing points, and the resulting WS wave functions are also localized similar to single-band approximation (\ref{WSsingle}). Such wave functions are given by Eqs.\ (\ref{A2})-(\ref{phiCx}).

\begin{figure}
\begin{center}\includegraphics[width=0.45\textwidth]{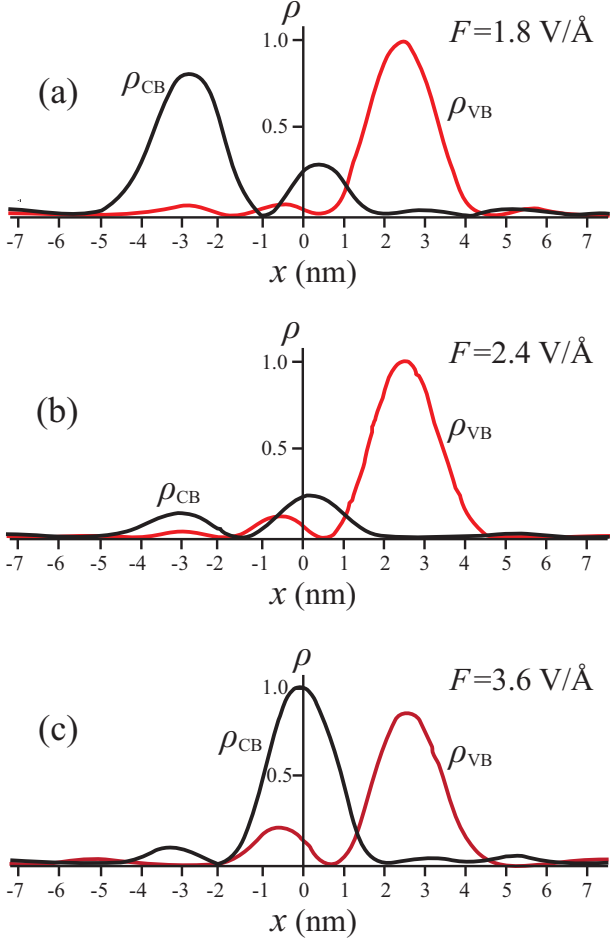}\end{center}
\caption{Electron densities $\rho_v(x)$ and $\rho_c(x)$ in the conduction and valence 
bands of a given WS state. The electric field is (a) $F=1.8$ V/\AA, (b) $F=2.4$ V/\AA, (c) $F=3.6$ V/\AA.  
The fields 1.8 V/\AA \ and 3.6 V/\AA \ corresponds to $l=2$ and $l=1$ anticrossing points. 
The $y$ component of the wave vector is $k_y = k_{y,0}$.  
} 
\label{WfunctionCBVB}
\end{figure}

To illustrate the interband mixing introduced by an electric field, we show in Fig.\ \ref{WfunctionCBVB} the
conduction and valence band probability densities for the WS wave functions, 
i.e., $\rho_v(x) = |\tilde{\phi}_v(x,k_y)|^2$ and $\rho_c(x) = |\tilde{\phi}_c(x,k_y)|^2$. The results 
are shown for one of the WS energy levels for a given electric field. 
The electric fields $F = 1.8 $ V/\AA \ and  $F = 3.6 $ V/\AA \ are near $l=2$ and $l=1$ anticrossing points, respectively. 
In these cases, the interband mixing is strong, and the electron densities in the conduction and valence bands are comparable [see Fig.\ \ref{WfunctionCBVB}(a), (c)]. The spatial separation between the maxima of $\rho_v(x)$ and $\rho_c(x)$ is $\approx la$. 
Thus for $F=1.8$ V/\AA, i.e., $l=2$, the distance between the maxima of $\rho_v$ and $\rho_c$ is $\approx 2a\approx 4.8$ \AA, while  for $F=3.6$ V/\AA, i.e., $l=1$, the distance
is $\approx a\approx 2.4$ \AA.

For electric field $F=2.4$ V/\AA, which is between $l=1$ and $l=2$ anticrossing points, the interband 
mixing is weak. In this case only one component (in our case only the valence band component, $\rho_v$) is strong [see Fig.\ \ref{WfunctionCBVB}(b)]. 

\begin{figure}
\begin{center}\includegraphics[width=0.4\textwidth]{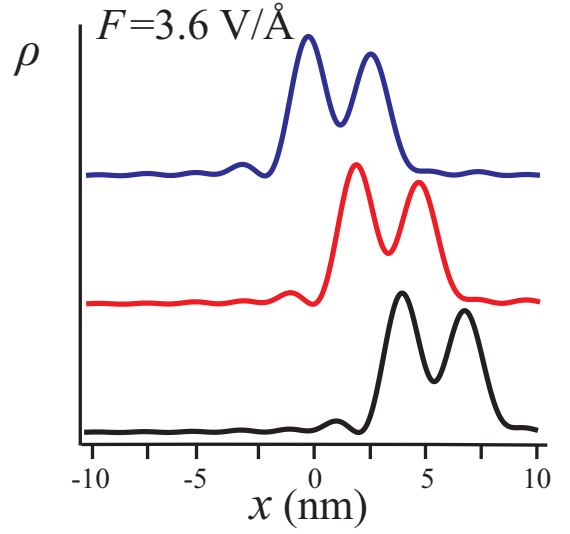}\end{center}
\caption{Total electron density $\rho(x) = \rho_v(x) + \rho_c(x)$ of three WS states. The electric field 
is $F=3.6$ V/\AA,  corresponding to the $l=1$ anticrossing point. 
The $y$ component of the wave vector is $k_y = k_{y,0}$. The curves are displaced vertically for clarity.
} 
\label{WFunctions36}
\end{figure}

In both cases, i.e., at the anticrossing points and away from them, the wave functions are localized in 
the $x$ space. The localization length depends on the electric field. The points, at which the WS wave functions 
are localized, depend on the energy of the WS states. In Fig.\ \ref{WFunctions36} the total electron density,
defined as $\rho(x) = \rho_v(x) + \rho_c(x)$, is shown for different WS states at electric field $F=3.6$ V/\AA, 
which correspond to $l=1$ anticrossing point. With changing the energy of the WS state, the electron 
density distribution is shifted as a whole along the $x$ axis.


\subsection{Wannier-Stark states of two-band model: numerical results}
\label{WS_2_band}

In the previous Section, analytical results for the WS spectra of the tight-binding model were obtained in the case of the $\delta$-function dipole matrix elements. Such strong dependence of the dipole matrix 
element on the wave vector occurs near the Dirac points. Away from the Dirac points, the  dipole matrix 
element $|D_x|$ as a function of the wave vector has a broad peak. In such a case, the WS energy spectra can be obtained numerically.

It is convenient to solve the system of the eigenvalue equations, 
(\ref{Eqs1})-(\ref{Eqs2}), by expanding functions 
$\phi_v(\mathbf{k})$ and $\phi_c(\mathbf{k})$ in terms of the WS 
wave function of individual bands, Eqs.\ (\ref{WSV1}) and 
(\ref{WSC1}), calculated without interband coupling. Thus 
\begin{eqnarray}
& & \phi _v (\mathbf{k}) = \sum_n {\cal A}_n \phi_{v,n}^{(0)}(\mathbf{k}) \label{phi_WS1}\\
& & \phi _c (\mathbf{k}) = \sum_n {\cal B}_n \phi_{c,n}^{(0)}(\mathbf{k}),  \label{phi_WS2}
\end{eqnarray} 
where index $n$ labels the WS states [see Eqs.\ (\ref{EWS}) and 
(\ref{EWScb})], ${\cal A}_n $ and ${\cal B}_n$ are the corresponding expansion 
coefficients. 
Substituting expressions (\ref{phi_WS1}) and (\ref{phi_WS2}) into 
Eqs.\ (\ref{Eqs1})-(\ref{Eqs2}), we obtain the 
system of eigenvalue equations on expansion coefficients ${\cal A}_n $ and ${\cal B}_n$,
\begin{eqnarray}
& & E {\cal A}_n =  E^{WS}_{v,n}  {\cal A}_n + 
F \sum_m {\cal D}_{nm} {\cal B}_n  \label{AB1} \\
& & E {\cal B}_n =  E^{WS}_{c,n}  {\cal B}_n + 
F \sum_m {\cal D}^{*}_{nm} {\cal A}_n,  \label{AB2}
\end{eqnarray}
where ${\cal D}_{nm}$ are dipole matrix elements, calculated 
between the WS wave functions of individual bands,
\begin{eqnarray}
& & {\cal D}_{nm} =   \left\langle \phi_{c,n}^{(0)} \right| 
   D_x(\mathbf{k}) \left|\phi_{c,n}^{(0)}  \right\rangle = 
   \nonumber \\
 & & \frac{1}{2k_0} \int_{-k_0}^{k_0} 
    dk_x   D_x(k_x,k_y)  \exp \Bigg[  \frac{i}{eF} \Bigg( 
 2 \int_{-k_0}^{k_x} E_{c}(k^{\prime },k_y) dk^{\prime }   
\nonumber \\
&& + (E^{WS}_{c,n}-E^{WS}_{v,m})(k_x+k_0)
  \Bigg) \Bigg]    
\end{eqnarray}

\begin{figure}
\begin{center}\includegraphics[width=0.45\textwidth]{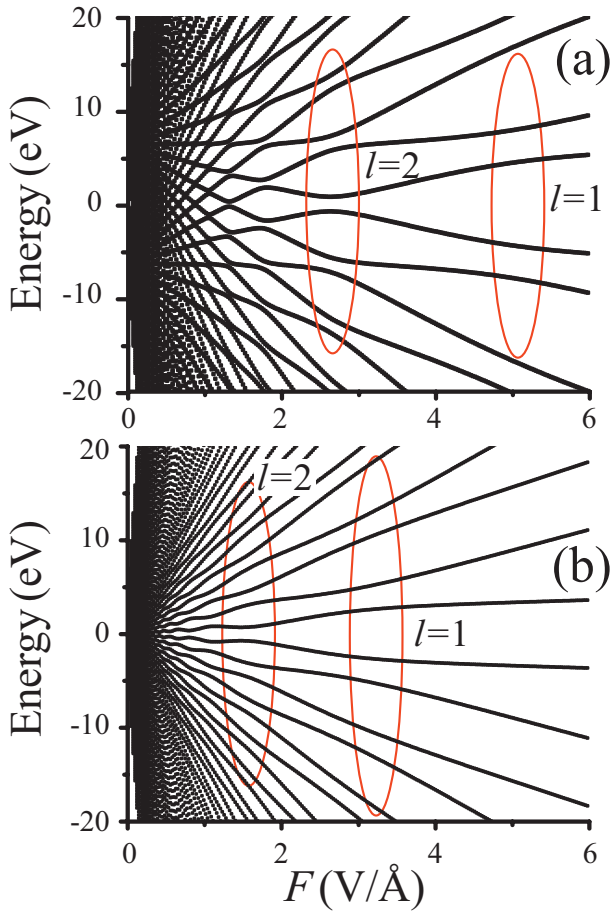}\end{center}
\caption{Energy spectra of graphene in a constant electric field, parallel to the $x$ axis. The spectra are 
calculated numerically for a finite size system for two values of $k_y$: (a) $k_y = 0$ and (b) $k_y = 0.32 (2\pi/a)$. 
 The number of states in each band is 100. The anticrossing points corresponding to $l=1$ and $l=2$ are marked by red lines. 
} 
\label{FigWS}
\end{figure}

In Fig.\ \ref{FigWS}, the energy spectra of a finite size system of graphene, calculated numerically from the system of equations (\ref{phi_WS1})-(\ref{phi_WS2}), are shown for different value of the $y$ component of the 
wave vector, $k_y$. At $k_y=0$ [see Fig.\ 
\ref{FigWS}(a)], the system is far away from the Dirac points. In this case, the dipole matrix element as a function of 
$k_x$ has a broad peak [see Fig.\ \ref{FigZ}]. For $k_y = 0.32 (2\pi/a)$ [see Fig.\ \ref{FigWS}(b)], the system is close to the 
Dirac point with the dipole matrix element having a sharp narrow peak. In this case, the values of 
the anticrossing gaps and the positions of the anticrossing points are close to the analytical expressions (\ref{FexactGr}) and 
(\ref{GapexactGr}), obtained in the model with $\delta$-function profile for the dipole matrix element.

The data, shown in Fig.\ \ref{FigWS}, illustrate strong dependence of the spectra on the value of $k_y$, i.e. on the shape of 
the function $D_x(k_x)$. With increasing $k_y\to K_y$, i.e. when the peak in $D_x(k_x)$ becomes sharp, the anticrossing points
move to smaller values of electric field and the anticrossing gaps become smaller.  

\begin{figure}
\begin{center}\includegraphics[width=0.47\textwidth]{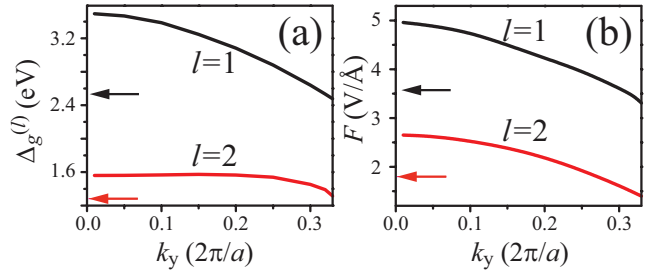}\end{center}
\caption{(a) Anticrossing gaps, calculated for the $l=2$ and $l=1$ anticrossing points, are shown as a function of the 
$y$ component of the wave vector, $k_y$. (b) The positions of $l=1$ and $l=2$ anticrossing points are shown as a function of 
$k_y$. The electric field is parallel to the $x$ axis. 
} 
\label{FigPositionGap2}
\end{figure}

In Fig.\ \ref{FigPositionGap2},
the anticrossing gaps and the positions of the anticrossing points are shown as a function of $k_y$ 
for $l=1$ and $l=2$ anticrossing points. 
A general trend is that with increasing $k_y$, both the anticrossing gaps, $\Delta_g^{(l)}$, 
and the electric fields, $F^{(l)}$, at which the anticrossing points are observed, are decreasing. The 
arrows in Fig.\ \ref{FigWS} show the analytical values of the anticrossing gaps and the
positions of the anticrossing points, obtained 
from Eqs.\ (\ref{FexactGr}) and (\ref{GapexactGr}). These numbers are close to the corresponding 
numerical values at $k_y \approx K_y= (1/3)(2\pi/a)$, i.e. near the Dirac point [see Fig.\ \ref{FigWS}].

\subsection{Wannier-Stark states of two-band model: two Dirac points} 

\begin{figure}
\begin{center}\includegraphics[width=0.45\textwidth]{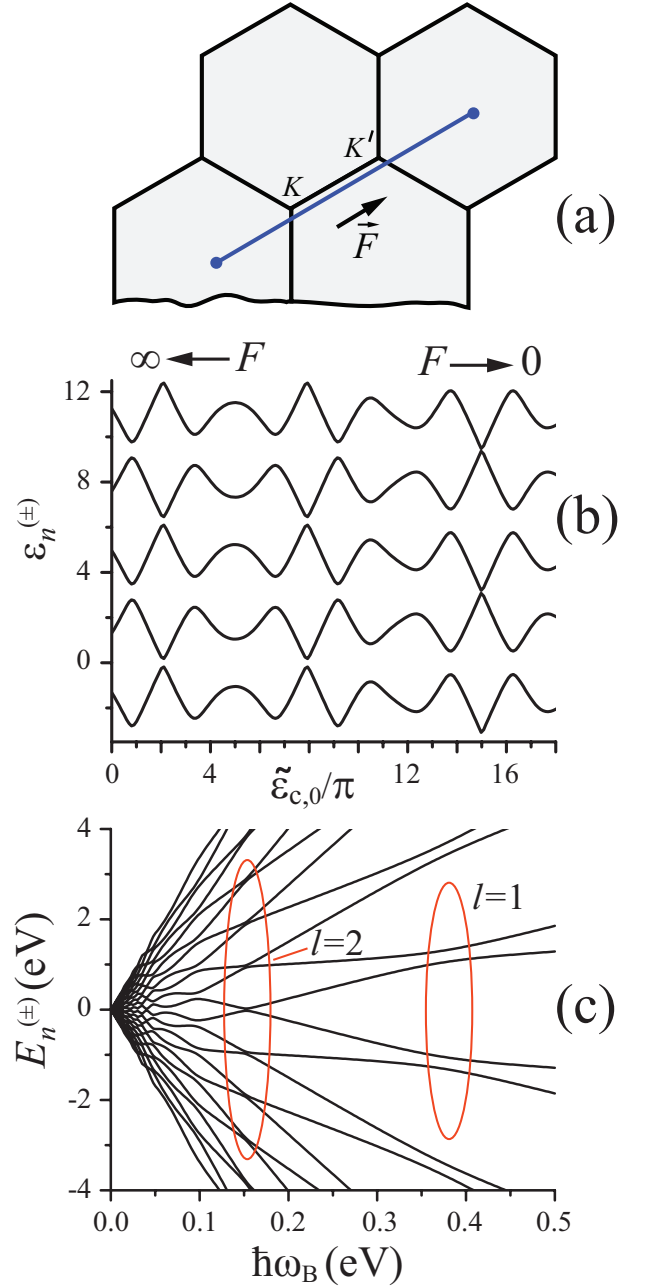}\end{center}
\caption{(a) Line of coupled states in the reciprocal space is shown by blue solid line. Along this line there are two 
inequivalent Dirac points $K$ and $K^{\prime }$. The direction of electric field is also shown. 
(b) Dimensionless energies $\varepsilon_n^{(\pm)}$ of WS states, calculated from 
Eq.\ (\ref{EnWS2DP2}), 
are shown as a function of dimensionless parameter $\tilde{\varepsilon}_{c,0}$ for different values of integer number $n$. The 
parameters 
$\Lambda_1 = \Lambda_2 = \Lambda_0$ and $\alpha$ are $\Lambda_0=0.6$ and $\alpha = 0.7$. 
(c) The energies $E_n^{(\pm)}$ of the WS states, calculated from Eq.\ (\ref{EnWS2DP}), are shown as a function 
of Bloch frequency, $\hbar \omega_B$, which is proportional to electric field. The anticrossing points, corresponding 
to $l=1$ and $l=2$, are marked by red lines. The parameters 
 are $\Lambda_0=0.6$, $\alpha = 0.7$, and $E_{c,0} =1$ eV. 
} 
\label{Fig2DP}
\end{figure}

By changing the direction of electric field, one can realize a situation when along a line of coupled states there 
are two Dirac points. For graphene, this happens for a line shown in Fig.\ \ref{Fig2DP}(a), i.e. when the angle between 
the direction of the electric field and axis $x$ is $\pi/6$. Then for the line, shown in Fig.\  \ref{Fig2DP}, we introduce 
one dimensional wave vector, $\kappa $, along the direction of electric field and write the dipole matrix element in terms of 
two $\delta$-functions, localized at the Dirac points, 
\begin{equation}
D_x(\kappa ) = e\Lambda_1 \delta(\kappa  - \kappa_1) + e\Lambda_2 \delta(\kappa  - \kappa_2),
\label{Ddelta2DP}
\end{equation}
where $\kappa_1$ and $\kappa_2$ are the coordinates of the Dirac points along the line of coupled states. 
The wave vector 
$\kappa$ changes from 0 to $\kappa_0 = (2\pi/a_{\kappa})$, where $a_{\kappa} = 2\pi/\kappa_0$ determines the period of the system along the 
direction of electric field.

We follow the same steps as in the case of one Dirac point (see Sec.\ \ref{Model1DP}). Namely, we introduce three regions, $0<\kappa < \kappa_1$, $\kappa_1<\kappa<\kappa_2$, and $\kappa_2<\kappa < \kappa_0$. In each region, the conduction and valence bands become
decoupled and the wave functions have the form of Eqs.\ (\ref{phiVD1})-(\ref{phiCD1}). At the boundary between the 
regions, i.e. at points $\kappa = \kappa_1$ and $\kappa = \kappa_2$, the boundary conditions have the form of Eqs.\ (\ref{relation1})-(\ref{relation2}). Combining all these equations and taking into account the periodic boundary conditions at 
poins $\kappa = 0$ and $\kappa= \kappa_0$, we obtain the following energy spectra of the coupled WS states
\begin{eqnarray}
E_n^{(\pm)} & = & \pm \frac{eF}{\kappa_0} \Bigg\{ 
\cos^{-1} \Bigg[  \cos \Lambda_1 \cos \Lambda_2  \cos \Bigg( \frac{\kappa_0}{eF}  \tilde{E}_{c,0}   \Bigg)  -   \nonumber \\
   & &  \sin \Lambda_1 \sin \Lambda_2  \cos \Bigg( \frac{\kappa_0}{eF}  \alpha \tilde{E}_{c,0}   \Bigg)
 \Bigg] + 2\pi n \Bigg\}.
\label{EnWS2DP}
\end{eqnarray}
Here $\tilde{E}_{c,0}$ is defined in terms of the linear integral over the line of coupled 
states (see Fig.\ \ref{Fig2DP}),
\begin{equation}
\tilde{E}_{c,0} = \frac{1}{\kappa_0} \int_0^{\kappa_0} E_c(\kappa ) d\kappa.
\end{equation}
The coefficient $0<\alpha <1$ in Eq.\ (\ref{EnWS2DP}) is defined by the following relation 
\begin{equation}
\alpha = 1 -  \frac{\textcolor{black}2}{\kappa_0 \tilde{E}_{c,0} } \int_{\kappa_1}^{\kappa_2} E_c(\kappa ) d\kappa.
\end{equation}
In dimensionless variables, $\varepsilon_n^{(\pm)} = E_n^{(\pm)} (\textcolor{black}{\kappa_0/eF})$ and 
$\tilde{\varepsilon}_{c,0} =  \tilde{E}_{c,0} (\kappa_0/eF)$, Eq.\ (\ref{EnWS2DP}) becomes
\begin{eqnarray}
\varepsilon_n^{(\pm)} & = & \pm \Bigg\{ 
\cos^{-1} \Bigg[  \cos \Lambda_1 \cos \Lambda_2  \cos \tilde{\varepsilon}_{c,0}  -   \nonumber \\
   & &  \sin \Lambda_1 \sin \Lambda_2  \cos ( \alpha \tilde{\varepsilon}_{c,0}   )
 \Bigg] + 2\pi n \Bigg\}.
\label{EnWS2DP2}
\end{eqnarray}

In Fig.\ \ref{Fig2DP}(b) the dimensionless WS energy spectrum (\ref{EnWS2DP2}) is shown 
for parameters $\Lambda_1 = \Lambda_2 = \Lambda_0=0.6$ and $\alpha = 0.7$, which \textcolor{black}{correspond to graphene}. 
Specific feature of this spectrum is a nonmonotonic dependence of the anticrossing gaps on the value of 
the dimensionless band offset, $\tilde{\varepsilon}_{c,0}$. These gaps have both large and very small values. The positions 
of the anticrossing points are also irregular. The corresponding energy spectrum in the original units is shown in 
Fig.\ \ref{Fig2DP}(c) as a function of electric field $F$. The anticrossing gaps have nonmonotonic dependence on $F$. 
For example, the anticrossing gap at $l=3$ is larger than the gap at $l=2$. This behavior is different from the 
behavior of the anticrossing gaps of the WS spectrum for systems where the dipole matrix elements are almost constant \cite{Apalkov_Stockman_PRB_2012_Strong_Field_Reflection}
or have a single peak as a function of the wave vector (see Sec.\ \ref{Model1DP}).

\section{Conclusion}

Within a single (either conduction or valence) band model, the energy spectrum of an electron in graphene in a constant external field
has a WS ladder structure with energy levels separated by the Bloch frequency, which is proportional to both the electric field and the 
lattice period of graphene crystal structure in the direction of electric field. In a two-band model, which is introduced above within the tight-binding nearest-neighbor approximation, a constant electric field results in
 mixing of the conduction and valence bands. As a result of such mixing, the energy spectrum of graphene as a function of electric field shows anticrossing points with the corresponding 
 anticrossing gaps. These gaps also indicate that a constant electric field opens a gap in the electron energy spectrum of graphene. This is understandable because it reduces symmetry of the system by lifting the equivalence (degeneracy) of the two constituent triangular sublattices.
The magnitudes of the gaps depend on the electric field. 
 
 The strength of the band mixing in an external electric field is determined by the magnitude of 
the interband dipole matrix element. The net (integral) interband dipole matrix element has a value of $-e\pi/3$ universally determined by the Pancharatnam-Berry phase.

In graphene, this interband dipole matrix element has unique dependence 
on the electron wave vector. Namely, at the Dirac points, it has sharp peaks, i.e., 
in the reciprocal space, the interband coupling is strong  near the Dirac points only. 
In this case, approximating such a strong  dependence of the dipole matrix element on the wave vector by the $\delta$-function, one can find an analytical expression for the WS energy spectrum. Such analytical solution predicts both the positions of the 
anticrossing points and the corresponding anticrossing gaps. 
As a function of inverse electric field the anticrossing points are equidistant. 
In the dimensionless units (relative to the Bloch frequency), 
the anticrossing gaps have the same value at all anticrossing points. Thus, in the original energy units, the anticrossing gaps are proportional to the electric field at the corresponding anticrossing points and, for graphene, 
are $\Delta_g^{(l)} = (2.54/l)$ (eV), where $l=1,2,\dots$ is an integer. Physically, such an anticrossing gap (divided by $\hbar$) is the rate of the Zener tunneling through $l$ unit cells that transfers an electron in a localized WS state from the valence to the conduction band.
 The largest anticrossing gap $\approx  2.54$ eV corresponds to the anticrossing point $l=1$ at the 
electric field  $\approx 3.59$ V/\AA. The weak dependence, $\propto l^{-1}$ of the anticrossing gaps on parameter $l$ is a
unique property of graphene and is due to highly nonuniform, singular profile of the dipole matrix element.

Such high fields, $F\gtrsim 1~\mathrm{V/\AA}$, can be generated only by laser pulses in the visible/near-infrared \cite{Schiffrin_at_al_Nature_2012_Current_in_Dielectric, Schultze_et_al_Nature_2012_Controlling_Dielectrics} or terahertz \cite{Huber_et_al_2014_nphoton_2013_349_HHG_in_GaSe} spectral regions.
Graphene  in a time-dependent electric field (see, for example, Ref.\ \cite{Avetissian_PRB_2013_Optical_Coherent_Control_in_Graphene}), when the electron dynamics is 
described in terms of the passage of the anticrossing points, the anticrossing gaps determine the 
characteristic time, $\tau_l = \hbar/ \Delta_g^{(l)} = 0.26 l$ fs, which characterizes adiabaticity of the dynamics. 
Namely, if time $\tau_p$ of passage of an anticrossing point, which is also the characteristic time of variation of
electric field, is much larger than $\tau_l$, $\tau_p\gg \tau_l$, then the electron dynamics is 
adiabatic. For example, if $\tau_l \approx 1$ fs, then the passages of anticrossing points $l=1$ and 2, which 
have the characteristic times  $\tau_l =0.25$ fs and 0.51 fs, are adiabatic, while the passages of the 
points $l>2$ is non-adiabatic or even diabatic. It is evident that no matter what is frequency range, from visible to terahertz, there always will be several anticrossings with near-resonant frequencies violating adiabaticity. Thus the rapid adiabatic passage \cite{Avetissian_PRB_2013_Optical_Coherent_Control_in_Graphene} is not possible in graphene; also Rabi oscillations will be dephased.

\section*{Appendix}
\label{Appendix}
 
 We express the general solution of the Schr\"odinger equation (\ref{HPsi}) - (\ref{Htotal}) in the form (\ref{PsiVC}), i.e., in the basis of eigenfunctions of field-free Hamiltonian ${\cal H}_0$. 
Substituting expression  (\ref{PsiVC}) for the wave function $ \Psi (\mathbf{r}) $ into the Schr\"{o}dinger equation (\ref{HPsi}) - (\ref{Htotal}), we obtain 
\begin{eqnarray}
& & E \sum_{\mathbf{k}_1} \left[\phi_v(\mathbf{k}_1) \Psi^{(v)}_{\mathbf{k}_1} (\mathbf{r}) +
                                     \phi_c(\mathbf{k}_1) \Psi^{(c)}_{\mathbf{k}_1} (\mathbf{r})      \right]  =  \left( {\cal H}_0 + e \mathbf{F} \mathbf{r} \right) \times
\nonumber \\
& &
\sum_{\mathbf{k}_1} \left[\phi_v(\mathbf{k}_1) \Psi^{(v)}_{\mathbf{k}_1} (\mathbf{r}) +
                                     \phi_c(\mathbf{k}_1) \Psi^{(c)}_{\mathbf{k}_1} (\mathbf{r})      \right]
\label{Aeq1}
\end{eqnarray}
We multiply both sides of Eq.\ (\ref{Aeq1}) by  $\Psi^{(v)*}_{\mathbf{k}} (\mathbf{r})$ and then integrate it by $ \mathbf{r}$. Taking into account that $\Psi^{(v)}_{\mathbf{k}} (\mathbf{r})$  are eigenfunctions of Hamiltonian ${\cal H}_0$, we obtain 
\begin{eqnarray}
& &  E   \phi_v(\mathbf{k})  =   
E_{v}(\mathbf{k})\phi_v(\mathbf{k})  +   \nonumber \\
& & e    \sum_{\mathbf{k}_1}  \phi_v(\mathbf{k}_1)   \int   d\mathbf{r}   \Psi^{(v)*}_{\mathbf{k}} (\mathbf{r})  (\mathbf{F} \mathbf{r})     \Psi^{(v)}_{\mathbf{k}_1} (\mathbf{r})   +  \nonumber \\
& & e    \sum_{\mathbf{k}_1}  \phi_c(\mathbf{k}_1)   \int   d\mathbf{r}   \Psi^{(v)*}_{\mathbf{k}} (\mathbf{r})  (\mathbf{F} \mathbf{r})     \Psi^{(c)}_{\mathbf{k}_1} (\mathbf{r})  .
\label{Aeq2}
\end{eqnarray}
Substituting explicit expression (\ref{functionC}) for   $\Psi^{(v)}_{\mathbf{k}} (\mathbf{r})$, we rewrite the second term in the right hand side of Eq.\ (\ref{Aeq2})  as follows
\begin{eqnarray}
& &   e  \sum_{\mathbf{k}_1}  \phi_v(\mathbf{k}_1)   \int   d\mathbf{r}   \Psi^{(v)*}_{\mathbf{k}} (\mathbf{r})  (\mathbf{F}    \mathbf{r} )    \Psi^{(v)}_{\mathbf{k}_1} (\mathbf{r})   =  \nonumber \\
& & \frac{e}{2}  \sum_{\mathbf{k}_1}  \phi_v(\mathbf{k}_1) \left( 1 + e^{i(\varphi_{k}-\varphi_{k_1})} \right)  
\int   d\mathbf{r}   (\mathbf{F}    \mathbf{r})  e^{i \mathbf{r}(\mathbf{k}-\mathbf{k}_1   ) } = \nonumber \\
& & \frac{e}{2}  \sum_{\mathbf{k}_1}  \phi_v(\mathbf{k}_1) \left( 1 + e^{i(\varphi_{k}-\varphi_{k_1})} \right)  
\left( -i\mathbf{F}    
\frac{\partial}{\partial\mathbf{k}_1}\right) \delta(\mathbf{k} - \mathbf{k}_1)  = 
\nonumber  \\
& & ie \mathbf{F} \frac{\partial\phi_v(\mathbf{k})}{\partial\mathbf{k}} + \frac{e}{2} \phi_v(\mathbf{k}) \mathbf{F}  
\frac{\partial\varphi_{k}}{\partial\mathbf{k}}  
,
\label{Aeq3}
\end{eqnarray}
where in the last line, in the sum (integral) over $\mathbf{k}_1$, we use integration by parts. The final expression 
contains an additional term $\frac{e}{2} \phi_v(\mathbf{k}) \mathbf{F}  
\frac{\partial\varphi_{k}}{\partial\mathbf{k}}$, which is not included in the system of equations (\ref{Eqs1})-(\ref{Eqs2})
since this term can be eliminated by substitution $\phi_v(\mathbf{k}) \rightarrow \phi_v(\mathbf{k}) e^{i(e/2) \varphi_{k} }$ and does not affect the energy spectrum of the system. 

The third term in the right hand side of Eq.\ (\ref{Aeq2}) can be rewritten as
\begin{eqnarray}
& &   e  \sum_{\mathbf{k}_1}  \phi_c(\mathbf{k}_1)   \int   d\mathbf{r}   \Psi^{(v)*}_{\mathbf{k}} (\mathbf{r})  (\mathbf{F}    \mathbf{r} )    \Psi^{(c)}_{\mathbf{k}_1} (\mathbf{r})   =  \nonumber \\
& & \frac{e}{2}  \sum_{\mathbf{k}_1}  \phi_c(\mathbf{k}_1) \left( -1 + e^{i(\varphi_{k}-\varphi_{k_1})} \right)  
\int   d\mathbf{r}   (\mathbf{F}    \mathbf{r})  e^{i \mathbf{r}(\mathbf{k}-\mathbf{k}_1   ) } = \nonumber \\
& & \frac{e}{2}  \sum_{\mathbf{k}_1}  \phi_c(\mathbf{k}_1) \left(- 1 + e^{i(\varphi_{k}-\varphi_{k_1})} \right)  
\left( -i\mathbf{F}    
\frac{\partial}{\partial\mathbf{k}_1}\right) \delta(\mathbf{k} - \mathbf{k}_1)  = 
\nonumber  \\
& & \frac{e}{2} \phi_c(\mathbf{k}) \mathbf{F}  
\frac{\partial\varphi_{k}}{\partial\mathbf{k}}   = \mathbf{F} \mathbf{D}(\mathbf{k} )  \phi_c(\mathbf{k}) ,
\label{Aeq4}
\end{eqnarray}
where the term proportional to
$\frac{\partial\phi_c(\mathbf{k})}{\partial\mathbf{k}}$ is zero due to orthogonality of the conduction and valence band 
free-field functions: 
\begin{equation}
\left(- 1 + e^{i(\varphi_{k}-\varphi_{k_1})} \right) \delta(\mathbf{k} - \mathbf{k}_1) = 0.
\end{equation}
Combining Eqs.\ (\ref{Aeq2})-(\ref{Aeq4}), we obtain Eq.\ (\ref{Eqs1}).
Similarly, multiplying Eq.\ (\ref{Aeq1}) by  $\Psi^{(c)*}_{\mathbf{k}} (\mathbf{r})$ and  integrating it by $ \mathbf{r}$, we can derive Eq.\ (\ref{Eqs2}).

\section*{Acknowledgment}

This work was supported by MURI grant N00014-13-1-0649 from the US Office of Naval Research, grant No. DE-FG02-11ER46789 from the Materials Sciences and Engineering Division, Office of the Basic Energy Sciences, Office of Science, U.S. Department of Energy, Grant No.\ DE-FG02-01ER15213 from the Chemical Sciences, Biosciences and Geosciences Division, Office of the Basic Energy Sciences, Office of Science, U.S. Department of Energy, and  NSF grant No.\ ECCS-1308473. 
MIS gratefully acknowledges also support by the Max Planck Society and the Deutsche Forschungsgemeinschaft Cluster of Excellence: Munich Center for Advanced Photonics (http://www.munich-photonics.de) during his Sabbaticals at Munich.


\end{document}